\documentstyle[12pt]{article}  
\begin{document}  \bibliographystyle{unsrt} \vbox{\vspace{30mm}}

\centerline{GROUP CONTRACTIONS: INONU, WIGNER, AND EINSTEIN}
\vspace{6mm}
\centerline{ Y.S.Kim\footnote{Internet: kim@umdhep.umd.edu}}
\centerline{\it Department of Physics, University of Maryland,}
\centerline{\it College Park, Maryland 20742, U.S.A.}
\vspace{10mm}

\abstract{Einstein's $E = mc^{2}$ unifies the momentum-energy relations
for massive and massless particles.  According to Wigner, the internal
space-time symmetries of massive and massless particles are isomorphic
to $O(3)$ and $E(2)$ respectively.  According to Inonu and Wigner, $O(3)$
can be contracted to $E(2)$ in the large-radius limit.  It is noted that
the $O(3)$-like little group for massive particles can be contracted to
the $E(2)$-like little group for massless particles in the limit of large
momentum and/or small mass.  It is thus shown that transverse rotational
degrees of freedom for massive particles become contracted to gauge
degrees of freedom for massless particles.}

\section{Introduction}\label{intro}
\vspace*{-0.5pt}
\noindent
In 1953 \cite{inonu53}, Inonu and Wigner published their classic paper
on group contractions.  The concept is very simple.  When we walk on
the surface of the earth, we are making Euclidean transformations
on a two-dimensional plane: translations and rotations around an
axis perpendicular to the plane.  Strictly speaking, however, we are
making rotations around the center of the earth with a very large
radius.  The work of Inonu and Wigner was based on this geometry.

In their paper, Inonu and Wigner were interested in unitary
representations of $E(2)$, and consequently they were interested in
mathematics in which the spherical harmonics become Bessel functions.
In this report, we point out that non-unitary finite dimensional
representations of $E(2)$ are also physically relevant and occupy an
important place in physics as well as in mathematics \cite{knp86}.
While Inonu and Wigner were interested only in the plane tangential to
the sphere with a large radius, we would like to point out here that
there is a cylinder tangential to the sphere.  We shall point out also
that the motion of a point on the cylindrical surface could be formulated
in terms of an $E(2)$-like group \cite{kiwi87jm,kiwi90jm,boya67}.

Let us now get to the main issue.  In 1939 \cite{wig39}, Wigner observed
that internal space-time symmetries of relativistic particles are
dictated by their respective little groups \cite{wig39}.  The little
group is the maximal subgroup of the Lorentz group which leaves the
four-momentum of the particle invariant.  He showed that the little
groups for massive and massless particles are isomorphic to the
three-dimensional rotation group and the two-dimensional Euclidean group
respectively.  Wigner's 1939 paper indeed gives a covariant picture
massive particles with spins, and connects the helicity of a massless
particle with the rotational degree of freedom in the group $E(2)$.
This paper also gives many homework problems for us to solve.  In this
report, let us concentrate ourselves on the following two questions.

\begin{itemize}
\item First, like the three-dimensional rotation group, $E(2)$ is a
three-parameter group.  It contains two translational degrees of freedom
in addition to the rotation.  What physics is associated with the
translational-like degrees of freedom for the case of the $E(2)$-like
little group?

\item Second, as is shown by Inonu and Wigner \cite{inonu53}, the
rotation group $O(3)$ can be contracted to $E(2)$.  Does this mean that
the $O(3)$-like little group can become the $E(2)$-like little group
in a certain limit?

\end{itemize}

As for the first question, it has been shown by various authors that
the translation-like degrees of freedom in the $E(2)$-like little group
is the gauge degree of freedom for massless particles \cite{janner71}.
As for the second question, it is not difficult to guess that the
$O(3)$-like little group becomes the $E(2)$-like little group in the
limit of large momentum/mass \cite{misra76}.  However, the non-trivial
result is that the transverse rotational degrees of freedom become
gauge degrees of freedom \cite{hks83}.  We can compare this result
to Einstein's energy-momentum relation $E = (m^{2} + p^{2})^{1/2}$,
which gives two different formulas for massive and massless particles.
The following table summarizes this comparison.

\vspace{6mm}
\begin{quote}Table I. Covariance of the energy-momentum relation, and
covariance of the internal space-time symmetry groups.\end{quote}
\vspace{5mm}
\begin{center}

\begin{tabular}{ccc}
Massive, Slow & COVARIANCE & Massless, Fast \\[2mm]\hline
{}&{}&{}\\
$E = p^{2}/2m$ & Einstein's $E = mc^{2}$ & $E = cp$ \\[4mm]\hline
{}&{}&{}  \\
$S_{3}$ & {}  &    $S_{3}$ \\ [-1mm]
{} & Wigner's Little Group & {} \\[-1mm]
$S_{1}, S_{2}$ & {} & Gauge Trans. \\[4mm]\hline

\end{tabular}

\end{center}

\vspace{6mm}

In Sec. 2, we shall discuss three-dimensional geometry of a sphere in
detail.  One of the three axes can become both contracted or expanded.
It is shown in  Sec. 3 that the two different deformations discussed
in Sec. 2 lead to Lorentz boost in the light-cone coordinate system.
The little groups are discussed in the light-cone coordinate system,
and the contraction of the $O(3)$-like little group to the $E(2)$-like
little group is calculated directly from the two different deformations
of the sphere.  In Sec. 4, we discuss possible applications of the
contraction procedure to other physical phenomena.

\section{Three-dimensional Geometry of the Little Groups}\label{3dim}
\vspace*{-0.5pt}
\noindent
The little groups for massive and massless particles are isomorphic
to $O(3)$ and $E(2)$ respectively.  It is not difficult to construct the
$O(3)$-like geometry of the little group for a massive particle at
rest \cite{wig39}.  The generators $L_{i}$ of the rotation group satisfy
the commutation relations:
\begin{equation}
[L_{i}, L_{j}] = i\epsilon _{ijk} L_{k} .
\end{equation}
Transformations applicable to the coordinate variables $x, y$, and $z$
are generated by
\begin{equation}\label{o3gen}
L_{1} = \pmatrix{0&0&0\cr0&0&-i\cr0&i&0} , \quad
L_{2} = \pmatrix{0&0&i\cr0&0&0\cr-i&0&0} , \quad
L_{3} = \pmatrix{0&-i&0\cr i &0&0\cr0&0&0} .
\end{equation}
The Euclidean group $E(2)$ is generated by $L_{3}, P_{1}$ and $P_{2}$,
with
\begin{equation}
P_{1} = \pmatrix{0&0&i\cr0&0&0\cr0&0&0} , \qquad
P_{2} = \pmatrix{0&0&0\cr0&0&i\cr0&0&0} ,
\end{equation}
and they satisfy the commutation relations:
\begin{equation}\label{e2com}
[P_{1}, P_{2}] = 0 , \qquad [L_{3}, P_{1}] = iP_{2} ,
\qquad [L_{3}, P_{2}] = -iP_{1} .
\end{equation}
The generator $L_{3}$ is given in Eq.(\ref{o3gen}).  When applied to the
vector space $(x, y, 1)$, $P_{1}$ and $P_{2}$ generate translations on in
the $x y$ plane.  The geometry of $E(2)$ is also quite familiar to us.

Let us transpose the above algebra.  Then $P_{1}$ and $P_{2}$ become
$Q_{1}$ and $Q_{2}$, where
\begin{equation}
Q_{1} = \pmatrix{0&0&0\cr0&0&0\cr i &0&0} , \qquad
Q_{2} = \pmatrix{0&0&0\cr0&0&0\cr0&i&0} ,
\end{equation}
respectively.  Together with $L_{3}$, these generators satisfy the
same set of commutation relations as that for
$L_{3}, P_{1}$, and $P_{2}$ given in Eq.(\ref{e2com})
\begin{equation}
[Q_{1}, Q_{2}] = 0 , \qquad [L_{3}, Q_{1}] = iQ_{2} , \qquad
[L_{3}, Q_{2}] = -iQ_{1} .
\end{equation}
These matrices generate transformations of a point on a circular cylinder.
Rotations around the cylindrical axis are generated by $L_{3}$.  The
$Q_{1}$ and $Q_{2}$ matrices generate the transformation:
\begin{equation}\label{cyltrans}
exp{\left(-i\xi Q_{1} - i\eta Q_{2}\right)} =
\pmatrix{1&0&0\cr0&1&0\cr \xi & \eta & 1} .
\end{equation}
When applied to the space $(x, y, z)$, this matrix changes the value of
$z$ while leaving the $x$ and $y$ variables invariant \cite{kiwi87jm}.
This corresponds to a translation along the cylindrical axis.  The
$J_{3}$ matrix generates rotations around the axis.  We shall call the
group generated by $J_{3}, Q_{1}$ and $Q_{2}$ the {\em cylindrical group}.

We can achieve the contractions to the Euclidean and cylindrical groups
by taking the large-radius limits of
\begin{equation}
P_{1} = {1\over R} B^{-1} L_{2} B ,
\qquad P_{2} = -{1\over R} B^{-1} L_{1} B ,
\end{equation}
and
\begin{equation}
Q_{1} = -{1\over R}B L_{2}B^{-1} , \qquad
Q_{2} = {1\over R} B L_{1} B^{-1} ,
\end{equation}
where
$$
B(R) = \pmatrix{1&0&0\cr0&1&0\cr0&0&R}  .
$$
The vector spaces to which the above generators are applicable are
$(x, y, z/R)$ and $(x, y, Rz)$ for the Euclidean and cylindrical groups
respectively.  They can be regarded as the north-pole and equatorial-belt
approximations of the spherical surface respectively.

Since $P_{1} (P_{2})$ commutes with $Q_{2} (Q_{1})$, we can consider the
following combination of generators.
\begin{equation}
F_{1} = P_{1} + Q_{1} , \qquad F_{2} = P_{2} + Q_{2} .
\end{equation}
Then these operators also satisfy the commutation relations:
\begin{equation}\label{commuf}
[F_{1}, F_{2}] = 0 , \qquad [L_{3}, F_{1}] = iF_{2} , \qquad
[L_{3}, F_{2}] = -iF_{1} .
\end{equation}
However, we cannot make this addition using the three-by-three matrices
for $P_{i}$ and $Q_{i}$ to construct three-by-three matrices for $F_{1}$
and $F_{2}$, because the vector spaces are different for the $P_{i}$ and
$Q_{i}$ representations.  We can accommodate this difference by creating
two different $z$ coordinates, one with a contracted $z$ and the other
with an expanded $z$, namely $(x, y, Rz, z/R)$.  Then the generators
become four-by-four matrices, and $F_{1}$ and $F_{2}$ take the form
\begin{equation}\label{f1f2}
F_{1} = \pmatrix{0&0&0&i\cr0&0&0&0\cr i &0&0&0\cr0&0&0&0} , \qquad
F_{2} = \pmatrix{0&0&0&0\cr0&0&0&i\cr0&i&0&0\cr0&0&0&0} .
\end{equation}
The rotation generator $L_{3}$ is also a four-by-four matrix:
\begin{equation}\label{2rot}
L_{3} = \pmatrix{0&-i&0&0\cr i&0&0&0\cr0&0&0&0\cr0&0&0&0} .
\end{equation}
These four-by-four matrices satisfy the $E(2)$-like commutation relations
of Eq.(\ref{commuf}).

Next, let us consider the transformation matrix generated by the above
matrices.  It is easy to visualize the transformations generated by
$P_{i}$ and $Q_{i}$.  It would be easy to visualize the transformation
generated by $F_{1}$ and $F_{2}$, if $P_{i}$ commuted with $Q_{i}$.
However, $P_{i}$ and $Q_{i}$ do not commute with each other, and the
transformation matrix takes a somewhat complicated form:
\begin{equation}\label{compli1}
\exp{\left\{-i(\xi F_{1} + \eta F_{2})\right\}} =
\pmatrix{1&0&0&\xi \cr0 & 1 & 0 & \eta
\cr \xi & \eta & 1 & (\xi ^{2} + \eta ^{2})/2 \cr0&0&0&1}  .
\end{equation}

\section{Little Groups in the Light-cone Coordinate System}\label{lcone}
\vspace*{-0.5pt}
\noindent
Let us now study the group of Lorentz transformations using the
light-cone coordinate system.  If the space-time coordinate is
specified by $(x, y, z, t)$, then the light-cone coordinate variables are
$(x, y, u, v)$ for a particle moving along the $z$ direction, where
\begin{equation}
u = (z + t)/\sqrt{2} , \qquad v = (z - t)/\sqrt{2} .
\end{equation}
The transformation from the conventional space-time coordinate to the
above system is achieved through a similarity transformation.

It is straight-forward to write the rotation generators $J_{i}$ and boost
generators $K_{i}$ in this light-cone coordinate system \cite{kiwi90jm}.
If a massive particle is at rest, its little group is generated by $J_{1},
J_{2}$ and $J_{3}$.  For a massless particle moving along the $z$ direction,
the little group is generated by $N_{1}, N_{2}$ and $J_{3}$, where
\begin{equation}
N_{1} = K_{1} - J_{2} , \qquad N_{2} = K_{2} + J_{1} ,
\end{equation}
which can be written in the matrix form as
\begin{equation}
N_{1} = {1\over\sqrt{2}}\pmatrix{0&0&0&i\cr0&0&0&0\cr i &0&0&0\cr0&0&0&0} ,
\qquad N_{2} = {1\over\sqrt{2}}\pmatrix{0&0&0&0\cr0&0&0&i\cr0&i&0&0\cr0&0&0&0} ,
\end{equation}
and $J_{3}$ takes the form of the four-by-four matrix given in Eq.(\ref{2rot})

These matrices satisfy the commutation relations:
\begin{equation}\label{e2comm2}
[J_{3}, N_{1}] =i N_{2} ,\qquad [J_{3}, N_{2}] = -i N_{1} , \qquad
[N_{1}, N_{2}] = 0 .
\end{equation}
Let us go back to $F_{1}$ and $F_{2}$ of Eq.(\ref{f1f2}).  Indeed, they
are proportional to $N_{1}$ and $N_{2}$ respectively.
Since $F_{1}$ and $F_{2}$ are somewhat simpler than $N_{1}$ and $N_{2}$,
and since the commutation relations of Eq.(\ref{e2comm2}) are invariant
under multiplication of $N_{1}$ and $N_{2}$ by constant factors, we shall
hereafter use $F_{1}$ and $F_{2}$ for $N_{1}$ and $N_{2}$.

In the light-cone coordinate system, the boost matrix takes the form
\begin{equation}\label{boost}
B(R) = \exp \pmatrix{-i\rho K_{3}} =
\pmatrix{1&0&0&0\cr0&1&0&0\cr0&0&R&0\cr0&0&0&1/R} ,
\end{equation}
with $\rho = \ln (R)$, and $R = \sqrt{(1 + \beta )/(1 - \beta)}$, where
$\beta$ is the velocity parameter of the particle.  The boost is along the
$z$ direction.  Under this transformation, $x$ and $y$ coordinates are
invariant, and the light-cone variables $u$ and $v$ are transformed as
\begin{equation}
u' = Ru , \qquad v' = v/R .
\end{equation}
If we boost $J_{2}$ and $J_{1}$ and multiply them by $\sqrt{2}/R$, as
$$
W_{1}(R) = -{\sqrt{2} \over R}BJ_{2}B^{-1} = \pmatrix{0&0&-i/R^{2}&i
\cr0&0&0&0 \cr i&0&0&0\cr i/R^{2}&0&0&0}  ,
$$
\begin{equation}\label{w1w2}
W_{2}(R) = {\sqrt{2} \over R} BJ_{1}B^{-1} = \pmatrix{0&0&0&0\cr0&0&-
i/R^{2}&i\cr0&i&0&0\cr0&i/R^{2}&0&0}  ,
\end{equation}
then $W_{1}(R)$ and $W_{2}(R)$ become $F_{1}$ and $F_{2}$ of
Eq.(\ref{f1f2}) respectively in the large-$R$ limit.

The most general form of the transformation matrix is
\begin{equation}
D(\xi, \eta, \alpha ) = D(\xi, \eta, 0)D(0, 0, \alpha) ,
\end{equation}
with
\begin{equation}
D(\xi,\eta,0) = \exp{\left\{-i(\xi F_{1} + \eta F_{2})\right\}} , \qquad
D(0,0,\alpha) = \exp{\left(-i\alpha J_{3}\right)} .
\end{equation}
The matrix $D(0, 0,\alpha)$ represents a rotation around the $z$ axis.
In the light-cone coordinate system, $D(\xi ,\eta ,0)$ takes the form of
Eq.(\ref{compli1}).  It is then possible to decompose it into
\begin{equation}
D(\xi, \eta, 0) = C(\xi, \eta) E(\xi, \eta) S(\xi, \eta) ,
\end{equation}
where
\begin{eqnarray}\label{ces}
S(\xi, \eta) &=& I + {1\over 2} \left[C(\xi, \eta), E(\xi, \eta)\right]
= \pmatrix{1&0&0&0 \cr 0&1&0&0 \cr 0&0&1&(\xi ^{2} + \eta ^{2})/2
\cr 0&0&0&1} ,\nonumber \\[2mm]
E(\xi, \eta) &=& \exp \pmatrix{-i\xi P_{1} - i\eta P_{2}} =
\pmatrix{1&0&0&\xi \cr0&1&0&\eta \cr0&0&1&0\cr0&0&0&1} , \nonumber \\[2mm]
C(\xi ,\eta ) &=& \exp \pmatrix{-i\xi Q_{1} - i\eta Q_{2}} =
\pmatrix{1&0&0&0 \cr 0&1&0&0 \cr \xi & \eta &1&0 \cr0&0&0&1} .
\end{eqnarray}

Let us consider the application of the above transformation matrix to
an electromagnetic four-potential of the form
\begin{equation}
A^{\mu }(x) = A^{\mu } e^{i(kz - \omega t)} ,
\end{equation}
with
\begin{equation}
A^{\mu} = \left(A_{1}, A_{2}, A_{u}, A_{v}\right) ,
\end{equation}
where $A_{u} = (A_{3} + A_{0})/\sqrt{2}$, and $A_{v} = (A_{3} -
A_{0})/\sqrt{2}$.  If we impose the Lorentz condition, the above
four-vector becomes
\begin{equation}
A^{\mu} = \left(A_{1}, A_{2}, A_{u}, 0\right) ,
\end{equation}
The matrix $S(\xi, \eta)$ leaves the above four-vector invariant.
The same is true for the $E(\xi, \eta)$ matrix.  Both $E(\xi, \eta)$
and $S(\xi, \eta)$ become identity matrices when applied to four-vectors
with vanishing fourth component.  Thus only the $C(\xi, \eta)$ matrix
performs non-trivial operations.  As in the case of Eq.(\ref{cyltrans}),
it performs transformations parallel to the cylindrical axis, which in
this case is the direction of the photon momentum.  It leaves the
transverse components of the four vector invariant, but changes the
longitudinal and time-like components at the same rate.  This is a
gauge transformation.

It is remarkable that the algebra of Lorentz transformations given in
this section can be explained in terms of the geometry of deformed
spheres developed in Sec. 2.

\section{Outlook}\label{concl}
\vspace*{-0.5pt}
\noindent
In this report, we discussed the group contraction procedure initiated
by Inonu and Wigner in 1953.  We have seen how this contraction
procedure can be applied to the question of unifying the internal
space-time symmetries of massive and massless particles.  The result
is summarized in Table I given in Sect. 1.

Indeed, many papers have been published on this subject since then,
and it is likely that there will be many more in the future.  This is
the reason why we are having this conference.  Algebraically speaking,
the contraction is the process of obtaining one set of Lie algebra
to another set through a limiting process.  Geometrically speaking,
the contraction is a transition from one surface to another through
a tangential area.  We are of course talking about groups with rich
representations.  Indeed, the concept of group contraction generates
very rich mathematics.

The question then is whether there are other physical phenomena which
can be fit into the framework defined in Table I.
In 1955 \cite{hofsta55}, Hofstadter and MacAllister reported their
experimental result showing that the proton is not a point particle but
has a space-time extension.  In 1964 \cite{gell64}, Gell-Mann proposed
the quark model in which the proton is a bound state of the constituent
particles called quarks.  The bound states so constructed is now called
hadrons.

On the other hand, in 1969 \cite{fey69}, Feynman observed that the
proton coming from a high-energy accelerator consists of ``partons" with
properties quite different from those of the quarks inside the hadron.
These days, the partons are routinely regarded as the quarks.  However,
it is not trivial to show why they have thoroughly different properties.
In order to solve this puzzle, we need a model of covariant bound states
which can be Lorentz-transformed.

Even before Hofstadter discovered the non-zero size of the proton,
Yukawa formulated a relativistic theory of extended hadrons based on
the covariant harmonic oscillators in 1953 \cite{yukawa53b}.  Yukawa's
idea was later extended to explain the quark model by Feynman,
Kislinger, and Ravndal \cite{fkr71}.  Yukawa's initial formalism was
then extended to accommodate Wigner's $O(3)$-like little group for
massive particles \cite{kno79}.

If we use this covariant oscillator formalism, it is possible to show
that the quark and parton models are two different manifestations of the
same covariant entity \cite{kn77a}, just as in the case of Einstein's
energy-momentum relation and Wigner's little group.  The result is
illustrated in the following table.

\vspace{6mm}
\begin{quote} Table II.  Addition of the quark-parton covariance to the
space-time symmetry table of Table I. \end{quote}
\vspace{5mm}
\begin{center}

\begin{tabular}{ccc}
Massive, Slow & COVARIANCE & Massless, Fast \\[2mm]\hline
{}&{}&{}\\
$E = p^{2}/2m$ & Einstein's $E = mc^{2}$ & $E = cp$ \\[4mm]\hline
{}&{}&{}  \\
$S_{3}$ & {}  &    $S_{3}$ \\ [-1mm]
{} & Wigner's Little Group & {} \\[-1mm]
$S_{1}, S_{2}$ & {} & Gauge Trans. \\[4mm]\hline
{} & {} & {}\\
Quark Model & Covariant Oscillators & Parton Model \\[5mm] \hline

\end{tabular}

\end{center}

\end{document}